%% file: main.tex
\documentclass[reprint,superscriptaddress,amsmath,amssymb, aps, prl, floatfix]{revtex4-2}

\usepackage{graphicx} 
\usepackage{amsmath,amsthm, amssymb}
\usepackage{bbm}
\usepackage{physics}
\usepackage{xcolor}
\usepackage{time}
\usepackage[pdftex,colorlinks=true]{hyperref}
\usepackage[colorinlistoftodos]{todonotes}
\usepackage{amsfonts}
\usepackage{mathtools}
\usepackage{bm}
\usepackage{comment}
\usepackage{ragged2e}
\usepackage{color,soul}
\usepackage{todonotes}
\newcounter{todocounter}

\def \be{\begin{equation}}
\def \ee{\end{equation}}
\def \bea{\begin{eqnarray}}
\def \eea{\end{eqnarray}}

\begin{document}

\title{Large Electron Model:  A Universal Ground State Predictor} 

\author{Timothy Zaklama}
\email{tzaklama@mit.edu}
\affiliation{Department of Physics, Massachusetts Institute of Technology, Cambridge, Massachusetts 02139, USA}

\author{Max Geier}
\affiliation{Department of Physics, Massachusetts Institute of Technology, Cambridge, Massachusetts 02139, USA}

\author{Liang Fu}
\email{liangfu@mit.edu}
\affiliation{Department of Physics, Massachusetts Institute of Technology, Cambridge, Massachusetts 02139, USA}

\begin{abstract}
We introduce \textit{Large Electron Model}, a single neural network model that produces variational wavefunctions of interacting electrons over the entire Hamiltonian parameter manifold. 
Our model employs the Fermi Sets architecture~\cite{fu2026universal}, a universal representation of many-body fermionic wavefunctions, which is further conditioned on Hamiltonian parameters and particle number. For interacting electrons in a two-dimensional harmonic potential, a single model accurately predicts the ground state wavefunction while generalizing across unseen coupling strengths and particle-number sectors, producing both accurate real-space charge densities and ground state energies, even up to $50$ particles. Our results establish a foundation model method for material discovery that is grounded in the variational principle, while accurately treating strong electron correlation beyond the capacity of density functional theory. 

\end{abstract}

\maketitle

\section{Introduction}
A central goal of materials science is to predict new properties of matter before they are measured. If this goal were achieved at scale, it would change how we discover materials, design molecules, and engineer quantum devices. While trial and error dominates the domain, 
accurate prediction of candidate materials with targeted functionality could transform materials research.
The rise of artificial intelligence (AI)  
raises the question: can AI predict new, experimentally verifiable material properties?
So far, studies in materials through AI have largely relied on existing datasets of experimentally measured material properties. But the properties we ultimately care about---equation of state, transport, magnetism, superconductivity, chemical reactivity---are all determined by the dynamics of interacting electrons as governed by the physical laws of quantum mechanics.

Reliable prediction beyond what has already been measured therefore must come from solving many interacting electrons at scale. While first-principles electronic structure is the most promising route to reliable prediction, accurate methods remain expensive and system specific, which is where progress has been slow for decades. Density functional theory (DFT) has been the workhorse of electronic structure calculations because it is versatile and scalable, enabling routine calculations on systems with a large number of electrons. However, despite its widespread success,   
DFT is not variational, does not produce many-body ground state wavefunctions, and completely fails in strongly correlated materials such as fractional Chern insulators or strong-coupling superconductors.   
%

Recently, a new approach to solve the many-electron problem using fermionic neural networks (NNs) \cite{luo2019backflow, Choo2020May,PauliNet2020,Pfau2020Sep} has shown great promise. In particular, self-attention neural networks \cite{vonGlehn2022Nov, geier2025self} are shown to accurately represent a wide variety of interacting electron states, including strongly correlated, highly entangled, and topologically ordered quantum matter that cannot be treated by DFT or any mean-field method \cite{teng2025solving, li2025attention,  nazaryan2025finding, fadon2025extracting, abouelkomsan2025topological}. More recently, a {\it provably} universal Fermi network has been found, which is capable of approximating any continuous Fermi wavefunctions to arbitrary accuracy using only a small number of Slater determinants \cite{fu2026universal}. 

By optimizing the network parameters through Monte Carlo energy minimization, fermionic NNs are able to solve the many-electron Schr\"odinger equation variationally, and find an accurate ground state energy and full quantum many-body wavefunction,  
from which {\it all} static and thermodynamic properties can be determined.     
Despite achieving high accuracy on many atomic, molecular and solid-state systems, the prevalent workflow of neural-network variational Monte Carlo (NN--VMC) still optimizes a new network for each Hamiltonian and each parameter choice from scratch, without the ability to reuse computation across different systems or sizes. 

The missing piece is a foundation model approach that solves for the many-electron quantum wavefunction across a manifold of physical settings with one shared set of parameters \cite{Bommasani2021,scherbela2022solving,gao2022abinitio,gao2023samplingfree,Gao2023Jul,zhu2023hubbard,Scherbela2024transferable,gerard2024transferable,foster2025ab,Rende2025,zaklama2025attention}. Broadly speaking, foundation models are trained on diverse and vast data sets to learn to predict beyond the available data. Applied to the materials context, foundation models replace single-instance solvers with parameter-sharing networks that learn transferable structure and enable rapid prediction after training. In our many-electron setting, supervised training is not viable, as the important problems do not come with ground-truth wavefunctions or energies at large system size.
For solving the many-electron problem at scale, a viable foundation model must learn unsupervised from the governing physical principle itself.

In this work, we introduce \textit{Large Electron Model (LEM)}, a foundation model for quantum wavefunctions of many-electron systems in the continuum. LEM is a single parameter-sharing network that conditions on Hamiltonian parameters, such as 
interaction strength, and particle number, and directly predicts the ground-state wavefunction for Hamiltonians it never saw. The model is fully unsupervised, thus it requires no reference solutions, no labeled training data, and no precomputed datasets; it is optimized end-to-end by variational energy minimization. The model is thus a universal ground state predictor that is variational by construction, rivaling the universality of DFT yet surpassing its reach in regimes of strong correlations, and guaranteeing a more interpretable and holistic solution.

Our foundation model is built from a {\it universal} Fermi wavefunction ansatz, which factorizes the wavefunction into an antisymmetric single-particle component and a symmetric correlation component \cite{fu2025minimal,fu2026universal}. The antisymmetric head allows for the most general fermionic sign structures through a compact Slater determinant basis, while the symmetric head learns electron correlations through an attention-based network designed to model systems with global dependencies. This architecture, which is both physically motivated and provably universal, yields a lightweight, efficient implementation that is well suited to capturing many-electron ground states from weak to strong interaction regimes. 

We demonstrate the power of our foundation model on the quantum dot problem, with interacting electrons in a two-dimensional harmonic trap at variable interaction strengths $\lambda$, and particle numbers $N$. After optimization via unsupervised energy minimization across interaction regimes and Hilbert-space sectors in a single run, we showcase the model's ability to accurately predict the entire family of ground state wavefunctions, whose energy expectation values beat benchmarks set by standard methods.  
Further, we show the model's remarkable generalization to predict the entire ground state wavefunction across interaction strength, and, critically, particle-number sectors through real space charge density plots. We also highlight the model's scalability, generalizing across the full interaction manifold for as large as $N=50$ particles. Our results establish a promising path beyond DFT for material discovery based on a universal and variational wavefunction model that 
can operate at large system sizes, while achieving unprecedented accuracy in strongly correlated regimes where DFT 
breaks down.


\begin{figure}
    \centering
    \includegraphics[width=\linewidth]{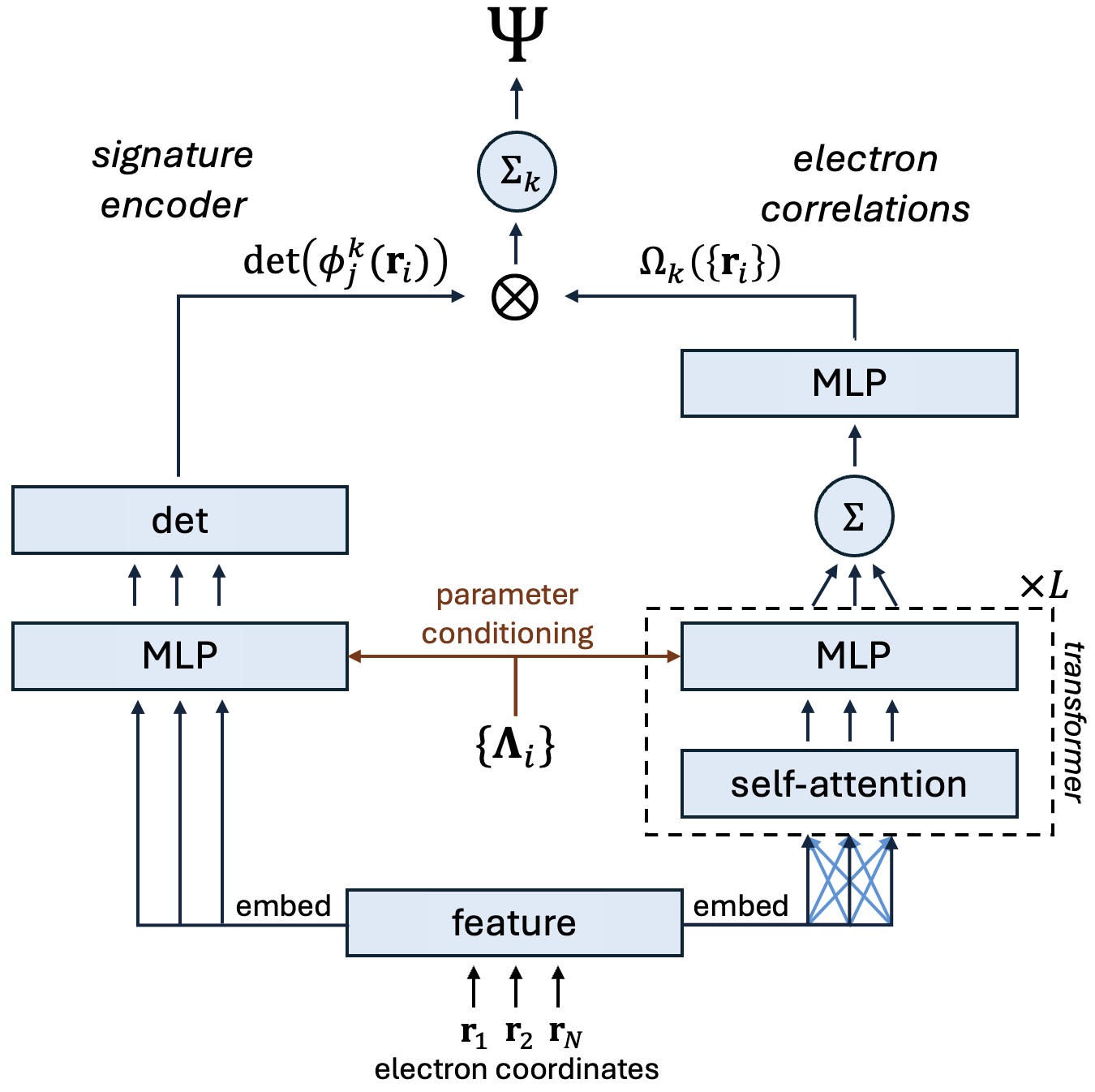}
    \caption{\textbf{Foundation model architecture.}
    A Multilayer Perceptron (MLP) with determinant head (left) constructs a set of $K$ signature encoders formulated as Slater determinants with orbitals represented by the MLP.
    Electron correlations are captured by a permutation-symmetric function represented by a transformer with symmetric head (right). 
    Conditioning both parts with the parameters $\bm \Lambda_i$
    yields the continuous parameterization of the many-body wavefunction $\Psi_{\bm \Lambda}(\{\bm r_i\})$. 
    }
    \label{fig:FermiSets}
\end{figure}

\section{Theoretical framework}
\label{sec:theory}

We consider a \emph{family} of continuum, many-electron Hamiltonians indexed by parameters
$\mathbf{\Lambda}\in\mathcal{P}$ (e.g., 
interaction strength and particle number).
For each $\mathbf{\Lambda}$, the ground state is a fermionic wavefunction in first quantization,
$\Psi_{\mathbf{\Lambda}}(\mathbf{R},\mathbf{s})$, defined on configurations
$\mathbf{R}=(\mathbf{r}_1,\dots,\mathbf{r}_N)\in(\mathbb{R}^d)^N$ with spin labels
$\mathbf{s}=(s_1,\dots,s_N)$.
Fermionic statistics impose antisymmetry under relabeling indistinguishable electrons:
\begin{equation}
\Psi_{\mathbf{\Lambda}}(\pi\!\cdot\!\mathbf{R},\pi\!\cdot\!\mathbf{s}) = (-1)^{\pi}\,\Psi_{\mathbf{\Lambda}}(\mathbf{R},\mathbf{s}),
\qquad \forall \pi\in S_N.
\end{equation}
Here, $S_N$ denotes the permutation group. 

Our goal is to learn a \emph{single} neural wavefunction $\Psi_{\theta}$ that represents the
entire manifold $\{\Psi_{\mathbf{\Lambda}}:\mathbf{\Lambda}\in\mathcal{P}\}$ of variational ground states by conditioning on $\mathbf{\Lambda}$:
\begin{equation}
\Psi_{\theta}:\; (\mathbf{R},\mathbf{s},\mathbf{\Lambda}) \mapsto \Psi_{\theta}(\mathbf{R},\mathbf{s};\mathbf{\Lambda})\in\mathbb{C},
\end{equation}
with shared parameters $\theta$ across all $\mathbf{\Lambda}$.
This ``parameter-sharing'' approach is the defining feature of the foundation model, since by optimizing on
an ensemble of Hamiltonians, the model learns a structured map $\mathbf{\Lambda}\mapsto \Psi_{\theta}(\mathbf{R},\mathbf{s};\mathbf{\Lambda})$
that can interpolate and generalize beyond the discrete optimization set. In our problem, the model must learn to interpolate in an infinite-dimensional space of $N$-particle wavefunctions in continuous space, a daunting task!   

We optimize $\Psi_{\theta}$ via the variational principle.
For each $\mathbf{\Lambda}$, define the normalized energy functional
\begin{equation}
E_{\theta}(\mathbf{\Lambda})
=
\frac{\langle \Psi_{\theta}(\mathbf{R},\mathbf{s};\mathbf{\Lambda}) \vert \hat{H}(\mathbf{\Lambda}) \vert \Psi_{\theta}(\mathbf{R},\mathbf{s};\mathbf{\Lambda}) \rangle}
{\langle \Psi_{\theta}(\mathbf{R},\mathbf{s};\mathbf{\Lambda}) \vert \Psi_{\theta}(\mathbf{R},\mathbf{s};\mathbf{\Lambda}) \rangle}.
\end{equation}
In variational Monte Carlo (VMC), this can be written as an expectation over the Born density
$\propto |\Psi_{\theta}(\mathbf{R},\mathbf{s};\mathbf{\Lambda})|^2$ using the local energy
$E_L(\mathbf{R},\mathbf{s};\mathbf{\Lambda})=\hat{H}(\mathbf{\Lambda})\Psi_{\theta}/\Psi_{\theta}$.
Given a finite optimization ensemble $\{\mathbf{\Lambda}_g\}_{g=1}^{G}\subset\mathcal{P}$, we minimize the
multi-task objective
\begin{align}
\mathcal{L}(\theta)
& =
\frac{1}{G}\sum_{g=1}^{G} E_{\theta}(\mathbf{\Lambda}_g) \\
& =
\frac{1}{G}\sum_{g=1}^{G}\;
\mathbb{E}_{(\mathbf{R},\mathbf{s})\sim |\Psi_{\theta}(\mathbf{R},\mathbf{s};\mathbf{\Lambda}_g)|^2}\!\left[
E_L(\mathbf{R},\mathbf{s};\mathbf{\Lambda}_g)
\right].
\label{eq:multitask_vmc}
\end{align}


\begin{figure}
    \centering
    \includegraphics[width=\linewidth]{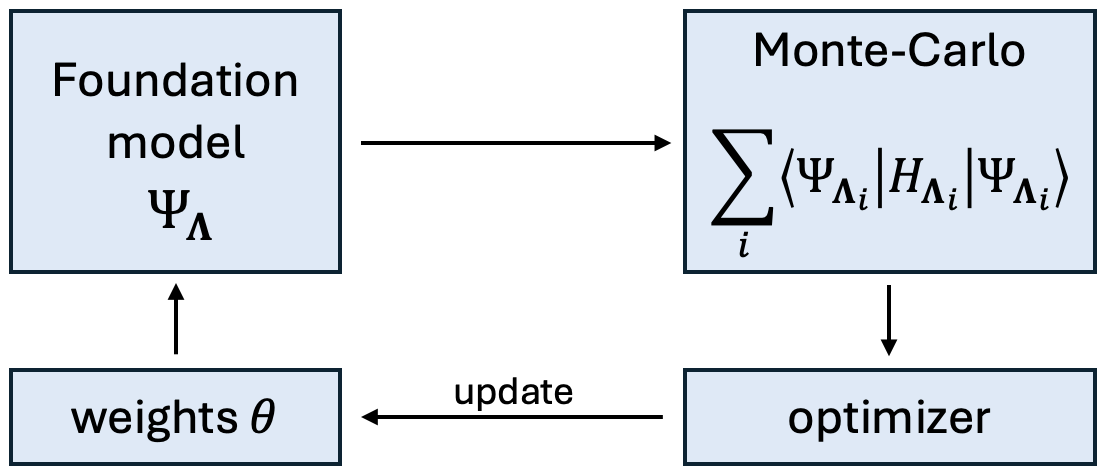}
    \caption{\textbf{Workflow.} The network represents the many-body wavefunction with continuous parameter dependence $\Psi_\mathbf{\Lambda}(\{ \mathbf{r}_i \})$. The loss is computed as the total energy $\sum_i \langle \Psi_{ \mathbf{\Lambda}_i}|H_{ \mathbf{\Lambda}_i}| \Psi_{ \mathbf{\Lambda}_i}\rangle $ accumulated over a fixed set of system parameters $\{ \mathbf{\Lambda}_i\}$, where the individual energy expectation values are computed using Monte-Carlo sampling. From the gradients of the total energy, an optimizer routine determines updates of network weights $\theta$.}
    \label{fig:workflow}
\end{figure}

\section{Approach}
\label{sec:approach}

We now introduce a universally representative foundation model for interacting fermions in the continuum.
The central novelty is a single conditional neural wavefunction that maps Hamiltonian parameters to ground-state
wavefunctions for real-space continuum systems, optimized once and reused across a manifold of physical settings.
To our knowledge, this is the first foundation-model formulation that targets real physical systems in the continuum via an explicitly fermionic, universally representative ansatz.

Let $N$ electrons in $d$ dimensions have configuration
$\mathbf{R}=(\mathbf{r}_1,\ldots,\mathbf{r}_N)$ with spins $\mathbf{s}=(s_1,\ldots,s_N)$.
Fermi Sets \cite{fu2026universal} expresses any continuous fermionic wavefunction as a sum of symmetric functions multiplying
antisymmetric cores,
\begin{equation}
\Psi_{\theta}(\mathbf{R},\mathbf{s};\mathbf{\Lambda})
=
\sum_{k=1}^{K}
\Omega_{\theta,k}(\mathbf{R},\mathbf{s};\mathbf{\Lambda})\;
\eta_{\theta,k}(\mathbf{R},\mathbf{s},\Lambda),
\label{eq:Fermi Sets_master}
\end{equation}
where each $\Omega_{\theta,k}$ $\in \mathbb{C}$ is invariant under any permutation of the electron labels, while each
$\eta_{\theta,k}$ is antisymmetric. This factorization guarantees fermionic antisymmetry, while allowing
the symmetric prefactors to adapt configuration-by-configuration. Here, we generalize the Fermi Sets to include parameter dependence for the purpose of building a foundation model. 

We begin by describing the construction of the symmetric component, $\Omega_{\theta,k}$.
We define geometric features relative to fixed reference centers $\{\mathbf{R}_A\}_{A=1}^{N_c}$
(e.g., the center of the harmonic trap or the position of a nucleus), to build particle features, 
\begin{align}
\mathbf{a}_{iA} &= \mathbf{r}_i-\mathbf{R}_A,\quad r_{iA}=\|\mathbf{a}_{iA}\|,
\end{align}
These coordinate features are concatenated with the electron spin to describe spinful wavefunctions \cite{Pfau2020,avdoshkin2025neural}. These features are then embedded into a latent space where they are processed by a transformer  \cite{Psiformer2022,geier2025self}. At this stage parameter conditioning is introduced to the parameter space $\mathbf{\Lambda}$ \cite{zaklama2025attention}. This operation $h_{\theta}$ on the particle features and parameter space yields the tokens
\begin{equation}
\mathbf{t}_i
=
h_{\theta}\!\Big(\{r_{iA},\mathbf{a}_{iA}\}_{A=1}^{N_c}, 
s_i;\mathbf{\Lambda}\Big)
\in\mathbb{R}^{f}.
\label{eq:electron_token}
\end{equation}
The output of the transformer is then pooled over the particle index to produce a permutation \emph{invariant} output,
\begin{equation}
\boldsymbol{\xi}_{\theta}(\mathbf{R},\mathbf{s};\mathbf{\Lambda})
=
\sum_{i=1}^{N} 
\sigma(\mathbf{t}_i),
\label{eq:xi_pool}
\end{equation}
where $\sigma$ is a nonlinearity,
ensuring $\boldsymbol{\xi}_{\theta}$ is invariant under electron permutation. For added expressivity, permutation-equivariant weights can be added as coefficients to each term in the sum.
The symmetric coefficients are then
produced by a small Multilayer Perceptron (MLP) 
$\rho_{\theta} : \mathbb{R}^{f} \to \mathbb{C}$:
\begin{equation}
\Omega_{\theta,k}(\mathbf{R},\mathbf{s};\mathbf{\Lambda})
=
\rho_{\theta}\!\big(\boldsymbol{\xi}_{\theta}(\mathbf{R},\mathbf{s};\mathbf{\Lambda})\big).
\label{eq:symmetric_coeffs}
\end{equation}

For the antisymmetric factor, we choose $\eta_{\theta,k}$ to be Slater determinants constructed from learned single-particle orbitals~\cite{geier2025self}:
\begin{align*}
\eta_{\theta,k}(\mathbf{R},\mathbf{s};\mathbf{\Lambda})
&=
\det \mathbf{M}^{(k)}_{\theta}(\mathbf{R},\mathbf{s};\mathbf{\Lambda}), \\
\big[\mathbf{M}^{(k)}_{\theta}(\mathbf{R},\mathbf{s};\mathbf{\Lambda})\big]_{ij}
&=
\phi^{(k)}_{\theta,j}(\mathbf{r}_i,s_i;\mathbf{\Lambda}),
\label{eq:slater_core}
\end{align*}
Antisymmetry is carried entirely by the Slater determinant.

Combining Eqs.~\eqref{eq:Fermi Sets_master}--\eqref{eq:symmetric_coeffs} yields the implemented conditional ansatz
\begin{equation}
\Psi_{\theta}(\mathbf{R},\mathbf{s};\mathbf{\Lambda})
=
\sum_{k=1}^{K}\rho_{\theta}\!\big(\boldsymbol{\xi}_{\theta}(\mathbf{R},\mathbf{s};\mathbf{\Lambda})\big)\;
\det \mathbf{M}^{(k)}_{\theta}(\mathbf{R},\mathbf{s};\mathbf{\Lambda}),
\label{eq:implemented_ansatz}
\end{equation}
which is shown schematically in Fig.~\ref{fig:FermiSets}.

\begin{table*}[ht]
\centering
\caption{Benchmarked energies using Fermi Sets ansatz for both single-run and foundation model against DMC and FermiNet references. The ``single task'' energies are produced after optimizing for a single system and sampling the converged trial wavefunction, whereas the foundation model (LEM) is optimized on many systems at once and then predicts the trial wavefunction, from which the energy is extracted after Monte Carlo sampling. The parentheses around the final digit denote the standard error on the previous digit. }
\label{tab:benchmarked-energies}
\begin{tabular}{c c c c c c}
\hline
$\lambda$ & $N$ & Fermi Sets (Single Task) & Fermi Sets (LEM) & DMC \cite{ghosal2007incipient} & FermiNet \cite{fadon2025interaction} \\
\hline
8.0  & 6 & $\mathbf{60.374(3)}$ & $\mathbf{60.3898(8)}$  & $60.3924(2)$   & $60.402(1)$  \\
10.0 & 6 & $\mathbf{68.93981(6)}$ & $\mathbf{68.942(1)}$           & $68.9458(4)$  & $68.967(1)$    \\
8.0  & 7 & $\mathbf{80.50917(8)}$ & $\mathbf{80.512(1)}$           & $80.5146(2)$  & $80.528(3)$  \\
8.0  & 8 & $\mathbf{103.0421(1)}$ & $\mathbf{103.043(3)}$           & $103.0464(3)$ & $103.0467(4)$  \\
\hline
\end{tabular}
\end{table*}

\section{Physical system} 
\label{sec:system}

To demonstrate the foundation model framework and its performance in the real world, we study the quantum dot problem. Explicitly, $N$ spin-polarized interacting electrons in $d=2$ are confined by an isotropic harmonic potential and interact through Coulomb repulsion.
Working in atomic units, we consider the standard many-electron Hamiltonian
\begin{equation}
\hat{H}(\mathbf{\Lambda})
=
-\frac{1}{2}\sum_{i=1}^{N}\nabla_i^2
+V_{\mathbf{\Lambda}}(\bm R),
\label{eq:harmonic_trap}
\end{equation}
with parametrized potential energy
\begin{equation}
V_{\mathbf{\Lambda}}(\bm R)
=
\frac{1}{2}\,\omega \sum_{i=1}^{N}\|\mathbf{r}_i\|^2
+\lambda \sum_{i<j}\frac{1}{\|\mathbf{r}_i-\mathbf{r}_j\|},
\end{equation}
and parameter vector $\mathbf{\Lambda}=(N,\omega,\lambda)$. Here $\omega$ sets the curvature of the parabolic confinement potential, and $\lambda$ controls the interaction strength. In all calculations in this work, we set $\omega=1$, which is equivalent to rescaling lengths by $\mathbf{x}_i=\omega^{1/4}\mathbf{r}_i$ and energies by
\begin{equation}
\hat{H}(N,\omega,\lambda)=\sqrt{\omega}\;\hat{H}\!\left(N,1,\frac{\lambda}{\omega^{1/4}}\right).
\end{equation}
Thus, results at $\omega\neq 1$ are obtained from the $\omega=1$ solutions through the single dimensionless coupling $\widetilde{\lambda}=\lambda/\omega^{1/4}$ and the overall energy scale $\sqrt{\omega}$.
 
Given $\Psi_{\theta}(\mathbf{R},\mathbf{s};\mathbf{\Lambda})$, the VMC local energy entering
Eq.~\eqref{eq:multitask_vmc} is
\begin{equation}
E_L(\mathbf{R},\mathbf{s};\mathbf{\Lambda})
=
\frac{\hat{H}(\mathbf{\Lambda})\,\Psi_{\theta}(\mathbf{R},\mathbf{s};\mathbf{\Lambda})}
{\Psi_{\theta}(\mathbf{R},\mathbf{s};\mathbf{\Lambda})}.
\end{equation}
We optimize on an ensemble of parameter points $\{\mathbf{\Lambda}_g\}_{g=1}^{G}$ spanning a target manifold,
thereby optimizing a single universal Fermi network that predicts ground-state wavefunctions across the family
$\{\hat{H}(\mathbf{\Lambda})\}_{\mathbf{\Lambda}\in\mathcal{P}}$. The workflow of the entire formalism can be visualized in Fig.~\ref{fig:workflow}.


\section{Results}
\label{sec:results}

We evaluate the model on interacting electrons in a two-dimensional harmonic trap across several interaction strengths, $\lambda$, and particle numbers, $N$. Once the model is optimized by minimizing the energy for a selected set of $\lambda$, $N$ parameters, it learns to predict the ground state wavefunction for any parameter set over the entire manifold spanned by the optimization set and even slightly beyond it. The ground state wavefunction is predicted instantaneously by the model, from which static and thermodynamic observables, such as the real space charge density and ground state energy, can be readily computed by Monte Carlo sampling. 
Here, we emphasize that no further optimization, transfer learning, or fine-tuning is performed; once our foundation model is optimized, it can predict many-electron ground states accurately, as we will show below.  

\begin{figure*}
    \centering
    \includegraphics[width=0.9\linewidth]{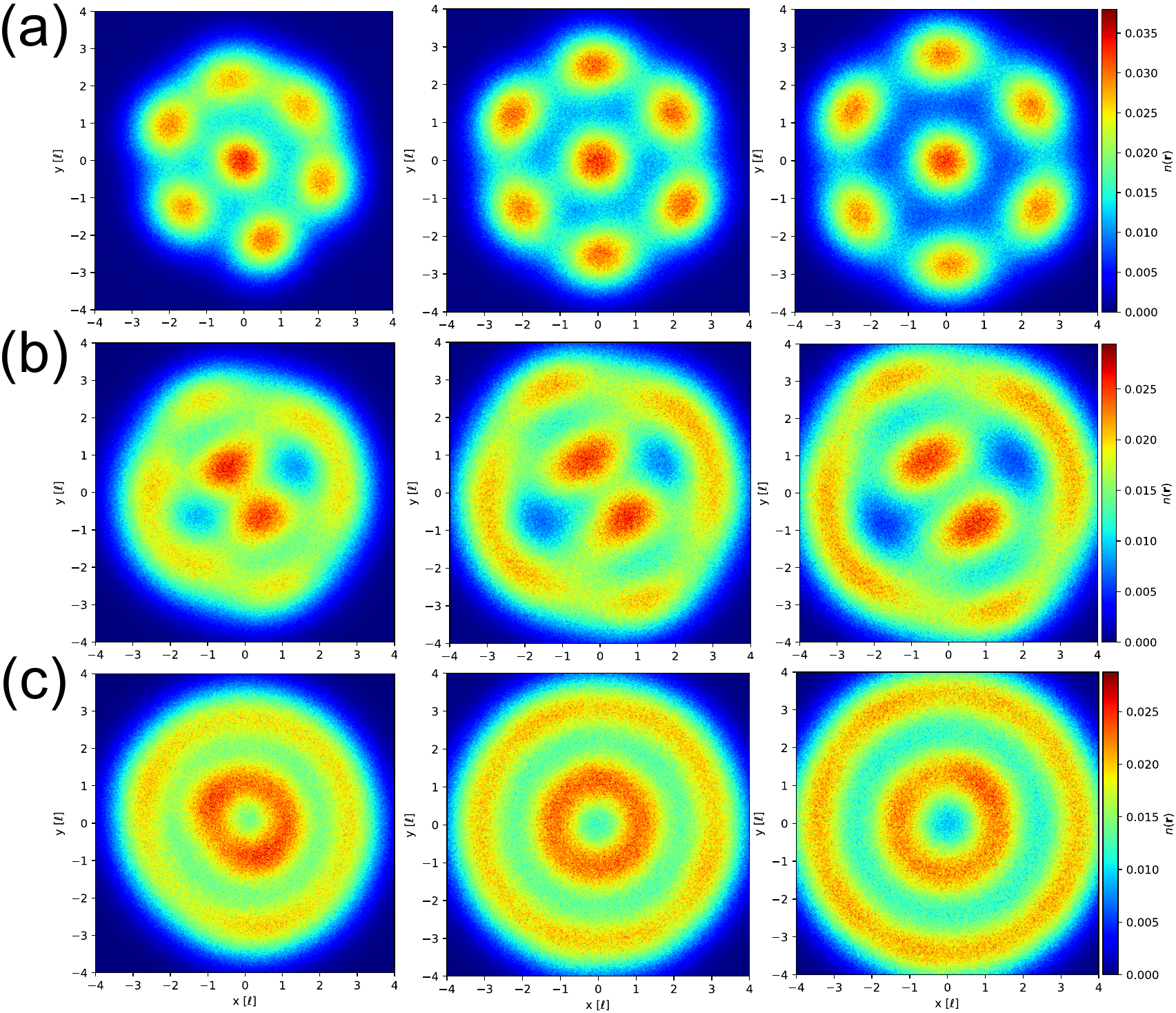}
    \caption{\textbf{Real space particle density from inference for various parameter values.} (a-c) are for $N=7,9,10$, respectively, with columns corresponding to $\lambda=3.0$ (predicted), $5.0$ (optimized for $N = 10$, predicted for $N = 7,9$), $7.0$ (predicted). Although the network is never optimized for any system with twofold angular structure, it is able to zero-shot predict the correct angular structure for the $N=9$ case.}
    \label{fig:densities}
\end{figure*}

The main result of this work is demonstrating that a single unified model can accurately predict wavefunctions for Hamiltonians it never saw. Optimizing only on a few interaction strengths $\lambda\in \{0.0,1.0,2.0,5.0,8.0,10.0\}$, and particle numbers $N\in \{6,8,10\}$, our model is able to accurately predict the ground state wavefunction throughout the continuous range $0\leq\lambda\leq10.0$, for every integer $5\leq N\leq11$. To demonstrate the accuracy, we compare with high-quality variational wavefunctions that are available for small particle numbers. 
We benchmark our energies against gold-standard Diffusion Monte Carlo (DMC) \cite{ghosal2007incipient} and FermiNet-based NN-VMC \cite{fadon2025interaction} for the quantum dot in Table~\ref{tab:benchmarked-energies}. 

From a single-run optimization on multiple systems, our foundational LEM beats DMC energies for all cases, while Fermi Sets individually optimized for each system reaches even lower energy. Importantly, the neural network produces an explicit many-electron ground state wavefunction, whereas DMC only represents the resulting ground-state distribution stochastically through walkers rather than as an explicit closed-form wavefunction. Thus, not only does our approach produce a more accurate ground state wavefunction, but it also allows any equal-time observable to be extracted directly. 
Importantly, our foundation model results are obtained without any problem-specific optimization or supervision. Rather, the model is optimized purely by unsupervised energy minimization and learns to predict ground states accurately and consistently across interaction strengths and particle numbers.  

To demonstrate the quality of the predicted many-body wavefunctions, we plot the predicted one-body density
\(n(\mathbf{r}) = N \int d\mathbf{r}_2\cdots d\mathbf{r}_N \sum_{\mathbf{s}} |\Psi(\mathbf{R},\mathbf{s})|^2\) in Fig.~\ref{fig:densities}. For $N=10$, 
the density remains rotationally symmetric for all \(\lambda\), as expected for a circular dot whose ground state is nondegenerate and has total angular momentum $L=0$ (``closed-shell''). Increasing interaction strength \(\lambda\) mainly
reshapes the radial profile as repulsion pushes charge density outward and sharpens correlation-driven radial structure,
producing the concentric ring features seen from \(\lambda=3\to 7\) \cite{ghosal2007incipient,reusch2001wigner,pederiva2000diffusion}. 

The open shell cases, \(N=7\) and \(N=9\), behave differently from the closed shell counterpart. For \(N=7\), the noninteracting ground state has a fourfold degeneracy with \(L=\pm 1,\pm 3\), arising from the $7$th electron occupying the fourth harmonic oscillator shell. At weak finite coupling, this degeneracy is lifted according to the ``modified Hund second rule'' for circular quantum dots, which favors occupation of the largest \(|l|\) orbitals within a partially filled shell after the spin sector has been fixed~\cite{ghosal2007incipient,fadon2025interaction}. Since the calculation is fully spin polarized, this rule selects the \(L=\pm 3\) manifold rather than the \(L=\pm 1\) manifold. Consistent with this expectation, the predicted densities at \(\lambda=3.0,5.0,7.0\) exhibit a sixfold angular structure, as expected for a linear combination of the \(L=+3\) and \(L=-3\) ground states that are exactly degenerate due to time-reversal symmetry. At stronger interaction, the same sixfold structure connects smoothly to the Wigner molecule regime, where seven repelling electrons form the classical \(1+6\) arrangement in a circular trap \cite{yannouleas1999spontaneous,rontani2006Full,reimann2002electronic,mitroy2013theory,sheikh2021symmetry}.  

The \(N=9\) case provides the particle hole counterpart within the same shell. Now, the modified Hund second rule first fills the \(l=+3\) and \(l=-3\) orbitals, leaving the remaining electron in either the \(l=+1\) or \(l=-1\) orbital. The predicted \(N=9\) densities thus correctly display the twofold structure, demonstrating the network's ability to predict the correct angular momentum sector $L=\pm 1$ despite never being optimized for a system with that angular momentum structure. We detail further the precise angular momentum decomposition of each wavefunction illustrated in Fig.~\ref{fig:densities} in the supplemental material~\cite{supplementary}.

\begin{figure}
    \centering
    \includegraphics[width=1\linewidth]{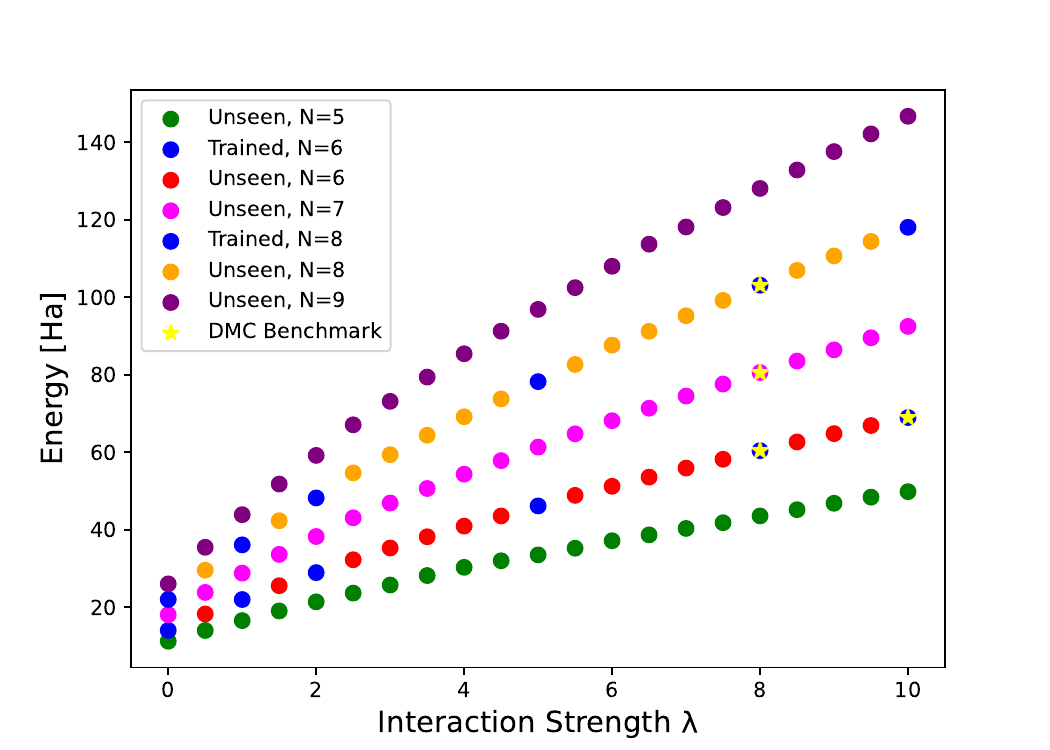}
    \caption{\textbf{Optimize and predict results across $(\lambda, N)$ particle grid.} {The model achieves pinpoint accuracy on out-of-sample results and can generate the wavefunction and any equal-time observable for any parameter across Hilbert-space sectors at inference time. Energies are competitive with state-of-the-art benchmarks \cite{ghosal2007incipient}, for which a detailed table is provided in the supplementary material~\cite{supplementary}. The statistical errors on the energies are far smaller than the radius of the dots, and therefore omitted in the figure.}}
    \label{fig:energy_VS_IS}
\end{figure}

Since \(n(\mathbf{r})\) is a
direct functional of \(|\Psi|^2\), these density plots are a stringent wavefunction-level check. Reproducing the correct symmetry patterns and correlation-driven features across both \(\lambda\) and \(N\) sectors provides evidence that the foundation model is learning the full landscape of ground-state wavefunctions in parameter space. 
Minor deviations do exist, which is expected when optimizing over energy alone, yet the qualitative and quantitative structure is faithful across unseen particle number and interaction strength, demonstrating the network's ability to predict wavefunction structure it had never learned directly, and highlighting the predictive power of the \textit{LEM}. With only a few instances during learning, it captures the correct symmetry physics and correlation-induced reorganization of the interacting ground states.

To demonstrate the model's accuracy over the entire parameter region, we plot the energy calculated from the predicted ground state wavefunction across the $(\lambda,N)$ grid. Fig.~\ref{fig:energy_VS_IS} shows that, after optimizing on a small set of parameters, the model produces pinpoint out-of-sample energies for unseen interaction strengths. More strikingly, the same learned parameters generalize across particle number as the model predicts accurate ground-state energies for completely unseen $N$ values. To our knowledge, this is the first neural quantum state that generalizes across distinct Hilbert-space sectors while still producing leading energy accuracy.
Previous foundation model studies on molecules and lattice spin models achieved accurate predictions for different Hamiltonian parameters \cite{gao2022abinitio,gao2023samplingfree,zhu2023hubbard,Rende2025}, but direct predictions for unseen system sizes were inaccurate and required further fine-tuning
\cite{scherbela2022solving,Scherbela2024transferable,gerard2024transferable,Gao2023Jul,foster2025ab}.
We emphasize that our model goes beyond an interpolation effect. All energies are computed from predicted ground states of $N$ particles, whose Hilbert space changes with $N$, yet the same foundation model produces accurate variational results. 

\begin{figure}
    \centering
    \includegraphics[width=0.95\linewidth]{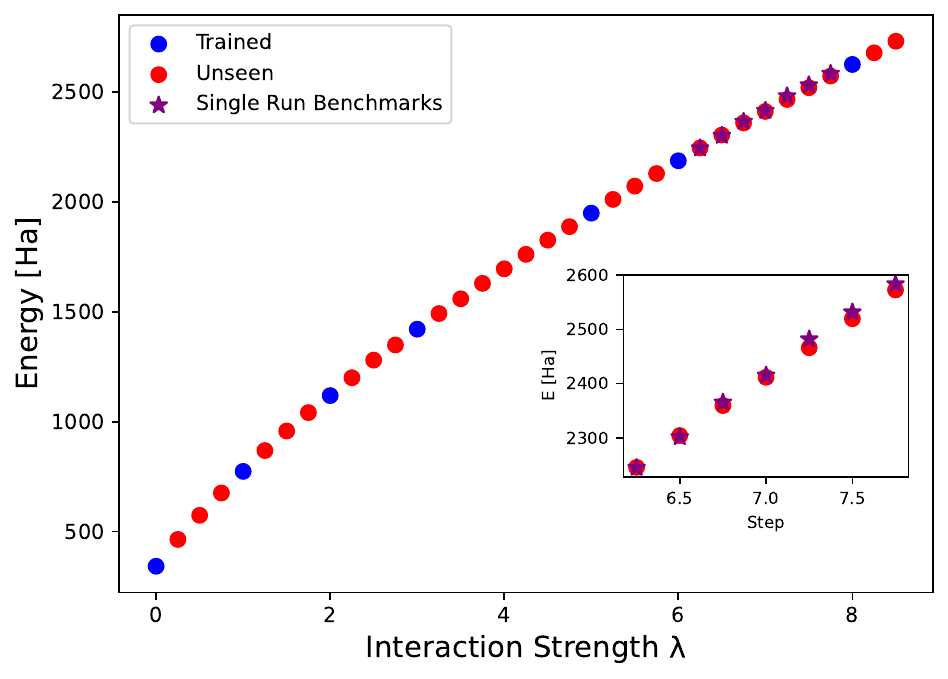}
    \caption{\textbf{Optimize and predict results for $N=50$ over the $\lambda$ parameter manifold.} {The model achieves pinpoint accuracy on out-of-sample results and can generate ground state energies that are even lower than single-parameter optimization and evaluation runs (highlighted in the inset). The statistical errors on the energies are far smaller than the radius of the dots, and therefore omitted in the figure.}}
    \label{fig:energy_VS_IS_50}
\end{figure}

Our approach also enables the model to efficiently predict the wavefunction over the entire family of coupling strengths for large $N$.
Again optimizing only on a few interaction strengths $\lambda\in \{0.0,1.0,2.0,3.0,5.0,6.0,8.0\}$, we find accurate predictions for unseen $\lambda$, at $N=50$, as illustrated in Fig.~\ref{fig:energy_VS_IS_50}, establishing a foundation model for many electrons in the continuum. In this regime, prediction is fast and low-cost as a single optimized model can predict across all interaction strengths without re-optimizing from scratch for each parameter point. In the majority of comparisons (for $6.0<\lambda<8.0$), the predicted out-of-sample runs outperform fresh single-system runs, indicating that parameter-conditioned, multi-task optimization improves the solution quality rather than diluting it. 
Optimizing on multiple related systems can act as an inductive bias that stabilizes optimization, flattens the loss landscape, and improves generalization.

While we demonstrate its performance on quantum dots, our \textit{LEM} is not tied to this particular system. Indeed, the architecture itself, based on Fermi Sets with parameter conditioning, is completely general and directly applicable to any electron systems, only requiring a specification of $\hat{H}(\mathbf{\Lambda})$ and a choice of conditioning parameters $\mathbf{\Lambda}$. The same model can be optimized in the same way on other many-electron settings, such as solids where electrons interact with the nuclei. Thus, these results establish a direct route to transferable ground state wavefunctions for real materials by building one trained model that can make accurate predictions across different materials and configurations. 


\section{Conclusion}
We introduced a foundation model for real physical systems in the continuum by constructing a single parameter-sharing neural network that maps system and Hamiltonian parameters directly to accurate many-body ground states. By combining a universal fermionic representation with parameter conditioning, the model learns a complete characterization of ground states across systems and Hamiltonian parameters. On interacting electrons in a two-dimensional harmonic trap, the model predicts the accurate many-body wavefunction and produces the state-of-the-art variational energies and expected charge density. 

The central advance is generalization at the wavefunction level. A single foundation model predicts accurate many-body ground states for interaction strengths it never saw, and, more importantly, for completely unseen particle numbers. This crosses Hilbert-space sectors where the input dimension and configuration space change, yet the model continues to output accurate Fermi wavefunctions. 
We further showcase the scaling of our model by generalizing over the entire interaction range, predicting ground state energies for up to $N=50$ electrons, and even predicting out-of-sample energies that outperform fresh single-system runs, demonstrating that parameter-conditioned multi-task optimization improves model capacity rather than diminishing it.


Because the method is fully unsupervised and requires only the Hamiltonian, it gives a direct route to foundation models for interacting electrons in the continuum, from quantum dots and trapped ions to atoms, molecules, and materials. 
The result is a powerful and precise formalism that can replace single-system workflows with one unified, reusable model, allowing for first-principles and large-scale studies of real materials.

\section{Acknowledgements}
We are grateful to Cyrus Umriger and Harold Baranger for constructive comments and illuminating insights. The authors acknowledge the MIT SuperCloud and Lincoln Laboratory Supercomputing Center for providing HPC resources that have contributed to the research results reported within this paper. This work made use of computing resources provided by the National Science Foundation under Cooperative Agreement PHY-2019786 (The NSF AI Institute for Artificial Intelligence and Fundamental Interactions).

\bibliography{ref}


\onecolumngrid
\newpage
\makeatletter 

\begin{center}
\textbf{\large Supplementary Material for \emph{Large Electron Model: A Universal Ground State Predictor}} \\[10pt]
Timothy Zaklama$^{1}$, Max Geier$^{1}$, and Liang Fu$^1$ \\
\textit{$^1$Department of Physics, Massachusetts Institute of Technology, Cambridge, MA-02139, USA}\\
\end{center}
\vspace{10pt}

\setcounter{figure}{0}
\setcounter{section}{0}
\setcounter{equation}{0}

\renewcommand{\thefigure}{S\@arabic\c@figure}
\makeatother

\appendix

\begin{center}
    \textbf{CONTENTS}
\end{center}

I. Table of benchmarked energies \hfill 2

II. Slater--Jastrow and Hartree--Fock comparison \hfill 2

III. Weak coupling density and angular momentum for open shell case \hfill 4

IV. Optimization stability across system size \hfill 6

V. Extended data \hfill 7

\include{supplement}

\end{document}

%% file: supplement.tex
\setcounter{equation}{0}
\setcounter{figure}{0}
\setcounter{table}{1}
\setcounter{page}{1}
\makeatletter
\renewcommand{\theequation}{S\arabic{equation}}
\renewcommand{\thefigure}{S\arabic{figure}}
\renewcommand{\citenumfont}[1]{S#1}
\renewcommand{\citenumfont}[1]{\textit{#1}}
\begin{widetext}

\section{Table of benchmarked energies}

\begin{table}[ht]
\centering
\caption{Ground-state energies for different benchmarked approaches at $N=10$. For the Neural Network Slater--Jastrow (NN--SJ) backflow, NN--SJ, and Hartree--Fock (HF) cases, energies are produced from calculations trained from scratch on a single system, using the same network size as the foundation model where applicable.
The training column denotes whether the Fermi Sets network saw that parameter value during training.}
\label{tab:numericalBenchmark}
\begin{tabular}{c c c c c c}
\hline
$\lambda$ & training & Fermi Sets (LEM) & NN--SJ backflow & NN--SJ & HF \\
\hline
1.0 & yes  & $\mathbf{52.0397}$  & $52.0559$  & $52.0796$  & $52.2161$   \\
3.0 & no & $\mathbf{87.5655}$  & $87.6543$  & $87.7070$          & $88.7643$   \\
5.0 & yes & $\mathbf{116.9503}$  &   $117.6311$      & $117.8350$  & $120.4366$   \\
7.0 & no & $\mathbf{142.9216}$    & $143.3966$ & $143.8019$  & $149.3424$   \\
\hline
\end{tabular}
\end{table}

We benchmark our energies against Neural Network (NN) Slater--Jastrow backflow and Hartree--Fock in Table~\ref{tab:numericalBenchmark}. With only a single training run, our predicted energies for both seen and unseen systems not only outperform these standard variational Monte Carlo techniques, but also produce a far more accurate ground-state wavefunction, as we describe below and in detail in the sections below. In all cases, the same network size is used, allowing the comparison to be most standard. Lower energies can be achieved by increasing the size of the network, as detailed further across the methods in the section below. It is important to note that the predicted energies from our \textit{Large Electron Model} are produced at once from a single run. In principle, the model can be further trained to optimize a single ground-state energy after training on many different parameter values, which would produce even lower energies. 

\section{Neural Network Slater--Jastrow and Hartree--Fock comparison}
\label{sec:supp_sj_hf}

Figure~\ref{fig:supp_densities} compares one-body densities at $\lambda=1.0,3.0,7.0$ obtained from two standard baselines: a single-determinant NN Slater--Jastrow trial state (top row) and Hartree--Fock (bottom row). For open-shell density diagnostics, this comparison also motivates our use of an interacting point such as $\lambda=3.0$ rather than the exactly noninteracting limit: at $\lambda=0$, the density can depend on the chosen representative within a degenerate shell, whereas at finite coupling the interaction selects the physically relevant correlated state. For a circular quantum dot, the Hamiltonian is rotationally invariant, and thus it commutes with the total angular-momentum operator
\begin{equation}
\hat{L}_z = \sum_{i=1}^{N} \hat{\ell}_{z,i} = -i \sum_{i=1}^{N} \frac{\partial}{\partial \theta_i},
\qquad [\hat{H},\hat{L}_z]=0.
\end{equation}
Exact eigenstates may therefore be chosen with definite total angular momentum $L$, and the corresponding one-body density
\begin{equation}
n(\mathbf{r}) = \left\langle \Psi \left| \sum_{i=1}^{N}\delta(\mathbf{r}-\mathbf{r}_i)\right|\Psi \right\rangle
\end{equation}
is rotationally symmetric for a single $L$ sector. Angular modulation in $n(\mathbf{r})$ arises when the approximate state mixes angular-momentum sectors (or forms a real linear combination of near-degenerate $\pm L$ sectors) according to 
\begin{equation}
    n(\theta)=\frac{1}{2\pi}\Big(1+2|ab^*|\cos\big(2l\theta + \arg(ab^*)\big) \Big),
\end{equation}
for amplitudes $a,b$ of nearly degenerate $\pm L$ sectors, which is precisely the failure mode highlighted by the Hartree--Fock densities at stronger coupling.

Hartree--Fock approximates the many-body state by a single Slater determinant,
\begin{equation}
\Psi_{\mathrm{HF}}(\mathbf{R},\mathbf{s})=\det\!\big[\chi_j(\mathbf{r}_i,s_i)\big],
\qquad
n_{\mathrm{HF}}(\mathbf{r})=\sum_{j=1}^{N}|\chi_j(\mathbf{r})|^2,
\end{equation}
and treats interactions at the mean-field exchange level. At weak coupling ($\lambda=1.0$), the density remains close to circular because the single-determinant picture is still reasonable. As $\lambda$ increases, correlations dominate and the mean-field solution lowers its energy by spontaneously localizing the orbitals, producing an azimuthally modulated, symmetry-broken charge density. At $\lambda=7.0$ the Hartree--Fock density becomes strongly non-uniform and irregular, reflecting that the self-consistent orbitals are not eigenstates of angular momentum but rather superpositions of many $m$ components; the determinant therefore mixes many total-$L$ sectors, and the one-body density no longer represents a controlled approximation to the rotationally invariant ground-state density. In this regime Hartree--Fock should be viewed as a broken-symmetry mean-field picture of localization rather than a faithful correlated eigenstate.

\begin{figure*}
    \centering
    \includegraphics[width=0.9\linewidth]{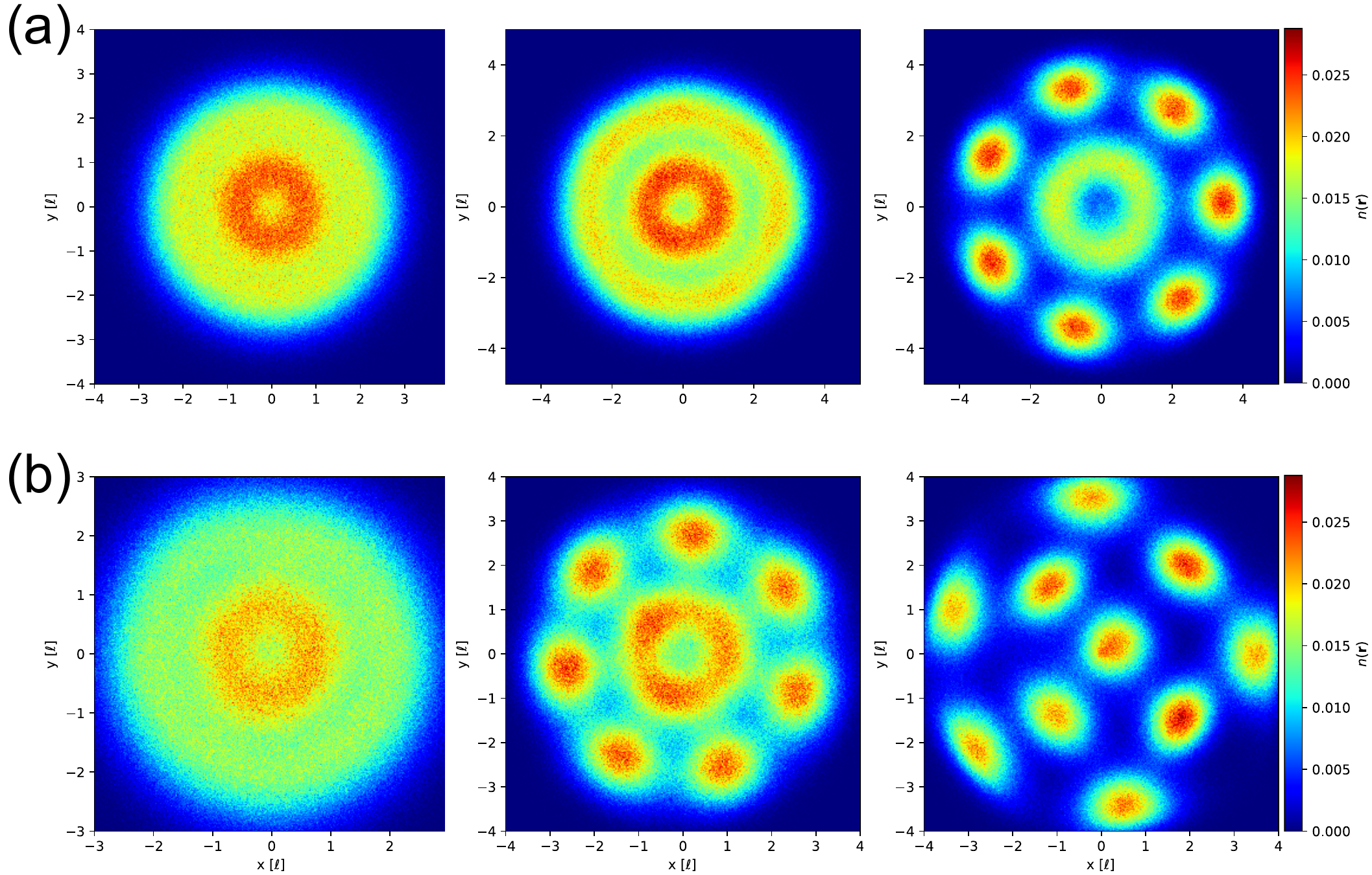}
    \caption{\textbf{Real-space particle density for NN Slater--Jastrow (a) and Hartree--Fock (b) calculations at $N=10, \lambda=1.0,3.0,7.0$.}}
    \label{fig:supp_densities}
\end{figure*}

The single-determinant NN Slater--Jastrow baseline partially repairs this by introducing explicit correlations on top of a determinant,
\begin{equation}
\Psi_{\mathrm{SJ}}(\mathbf{R},\mathbf{s})
=
\exp\!\Big(J(\mathbf{R})\Big)\,\det\!\big[\phi_j(\widetilde{\mathbf{r}}_i,s_i)\big],
\qquad
J(\mathbf{R})=\sum_{i<j}u(r_{ij})+\sum_i \nu(r_i)+\cdots,
\end{equation}
where the benchmark used here contains a single determinant. Multi-determinant Slater--Jastrow expansions, in which the determinant factor is replaced by a configuration expansion $\sum_{\alpha} c_{\alpha}D_{\alpha}$, have also become standard in modern quantum Monte Carlo and can substantially improve nodal flexibility. However, they introduce an additional multiconfigurational expansion and therefore go beyond the scope of this compact baseline comparison.

The Jastrow factor is taken to be real, $J(\mathbf{R})\in\mathbb{R}$, so $\exp(J(\mathbf{R}))$ is strictly positive for all configurations. This is the standard Slater--Jastrow convention: the Jastrow factor reshapes the amplitude and captures correlation effects, but by itself it does not change the antisymmetric sign structure or nodal surface inherited from the determinant.

In addition, our NN Jastrow is more expressive than a traditional low-parameter Jastrow ansatz. Standard Jastrow factors are often built from explicit one-body and two-body terms, such as $\sum_i \nu(r_i)$ and $\sum_{i<j}u(r_{ij})$. By contrast, the neural Jastrow used here is represented by multiple nonlinear layers acting on the full electronic configuration, so its effective expansion contains higher-order many-electron correlations beyond a simple pairwise form. Thus, even within a single-determinant Slater--Jastrow structure, this baseline is stronger than a conventional two-body Jastrow ansatz.

For the no-backflow version, $\widetilde{\mathbf{r}}_i=\mathbf{r}_i$; the same notation also allows the optional integration of a backflow component to further improve the nodal structure and recover additional correlation energy. The backflow coordinates $\widetilde{\mathbf{r}}_i$ are smooth, permutation-equivariant displacements of the bare electron positions that depend on the full configuration,
\begin{equation}
\widetilde{\mathbf{r}}_i
=
\mathbf{r}_i + \sum_{j\neq i}\eta(r_{ij})\,(\mathbf{r}_i-\mathbf{r}_j) + \cdots,
\end{equation}
(with $\eta$ a learned or parameterized scalar function).
The Jastrow factor captures the short-range correlation hole and a substantial fraction of dynamic correlation, while backflow improves the nodal structure by allowing the determinant to depend on correlation-distorted coordinates. This is reflected in Fig.~\ref{fig:supp_densities}: compared to Hartree--Fock, the single-determinant NN Slater--Jastrow (no backflow) densities remain closer to the expected circular profile at weak and intermediate coupling, and when angular features appear at strong coupling they do so in a more regular, physically interpretable pattern. However, even with backflow the ansatz remains limited by its restricted determinant backbone and cannot represent the full multiconfigurational mixing that becomes essential in the strongly correlated regime (also making it unstable and difficult to optimize at higher couplings). As a result, Slater--Jastrow can still exhibit residual angular-momentum mixing and does not match the wavefunction-level fidelity achieved by the foundation model.

In contrast, our Fermi Sets foundation model retains a compact antisymmetric core while learning correlations through a symmetric, configuration- and parameter-dependent mixing of antisymmetric components. This allows the model to represent correlation-driven reorganization without relying on uncontrolled symmetry breaking, and it produces densities that track the ground-state structure across $\lambda$ and across particle-number sectors. The comparison in Fig.~\ref{fig:supp_densities} therefore serves as a wavefunction-level diagnostic: Hartree--Fock breaks down first and most severely, the single-determinant NN Slater--Jastrow baseline captures a large portion of correlations but remains limited by its restricted antisymmetric backbone, while optional backflow can improve but not remove this structural constraint, and the Fermi Sets foundation model provides the most faithful densities consistent with its superior energy accuracy. Comparison of energies is shown in Table~\ref{tab:numericalBenchmark}.

\section{Weak coupling density and angular momentum for the open-shell case}
\label{sec:supp_n7_lambda1_angular_momentum}

The $N=7$ spin-polarized quantum dot provides a sensitive test of inference in an open shell. In the noninteracting limit, the first six electrons form a closed shell with total angular momentum $L=0$, while the seventh electron enters the next harmonic oscillator shell. This shell contains the degenerate single-particle orbitals
\begin{equation}
    (n,l)=(0,\pm 3),\qquad (n,l)=(1,\pm 1).
\end{equation}
The exact noninteracting ground state therefore contains both $|L|=3$ and $|L|=1$ sectors. At finite interaction strength, this degeneracy is lifted by the modified Hund second rule for circular quantum dots~\cite{ghosal2007incipient, fadon2025interaction}. Since the calculation is fully spin-polarized, the spin sector is fixed and the relevant choice is between the $|L|=3$ and $|L|=1$ orbital sectors. The weakly interacting ground state is expected to select the $|L|=3$ sector. The spin component of this shell filling logic is connected to Hund's first rule, whose microscopic origin in atoms and quantum dots has been analyzed in terms of exchange, electronic repulsion, and Fermi hole structure~\cite{sajeev2008hund,sako2012origin}.

The physical origin of this selection should not be interpreted as a simple preference for the most spatially extended orbital. Within this shell, the radially excited $(n,l)=(1,\pm 1)$ orbital has an outer radial lobe and carries more large radius weight than the nodeless $(n,l)=(0,\pm 3)$ orbital. The modified Hund second rule instead reflects the interaction-induced self-consistent potential. Coulomb repulsion flattens the parabolic confinement and makes the effective radial potential more hard-wall-like. In such a potential, the first radial mode with $|l|=3$ is lowered relative to the second radial mode with $|l|=1$, so the interacting weak-coupling state favors the $(0,\pm 3)$ pair and hence the $L=\pm 3$ manifold~\cite{ghosal2007incipient}.

\begin{figure*}
    \centering
    \includegraphics[width=0.9\linewidth]{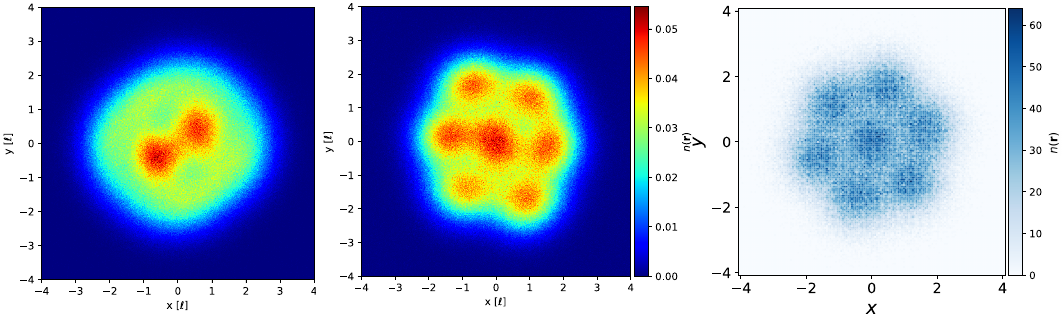}
    \caption{\textbf{One-body density plots for the $N=7$, $\lambda=1.0$ system.} Left: Density predicted by the same model as in the main text (training set $\lambda\in \{0.0,1.0,2.0,5.0,8.0,10.0\}$ and $N\in\{6,8,10\}$). The density exhibits a two-lobed angular structure, indicating that the inferred state is not the expected weak-coupling $|L|=3$ representative. Middle: Density predicted by a model with the same network size and training points as in the main text, but with $\lambda=10$ replaced by $\lambda=0.5$ to focus more on weak coupling (training set $\lambda\in \{0.0,0.5,1.0,2.0,5.0,8.0\}$ and $N\in\{6,8,10\}$). Right: Density of wavefunction optimized for the single $N=7$, $\lambda=1.0$ system (Single Fermi Sets optimization, no foundation model). The corresponding energy is 28.7929(1). Both the model with an additional weak-coupling training point and the single-system optimization predict the correct sixfold pattern evincing the $|L|=3$ representative.}
    \label{fig:supp_n7_lam1_L_projection}
\end{figure*}

Figure~\ref{fig:supp_n7_lam1_L_projection} shows how this open-shell diagnostic depends on the training distribution. The model used in the main text was trained on $N\in\{6,8,10\}$ and $\lambda\in\{0.0,1.0,2.0,5.0,8.0,10.0\}$. When evaluated at the unseen particle number $N=7$ and $\lambda=1.0$, it produced a twofold density pattern, indicating that the inferred state had selected the wrong open-shell sector. We then trained a model with the same architecture and the same number of training points, but rebalanced the interaction grid toward weak coupling by replacing $\lambda=10.0$ with $\lambda=0.5$. This modified training set produces the correct sixfold density at $N=7$, $\lambda=1.0$, in agreement with an independently optimized single-system Fermi Sets calculation.

This comparison demonstrates that the previous twofold result was not a physical transition and was not a structural limitation of the ansatz. It was an inference error caused by an insufficiently resolved weak-coupling training distribution near a delicate open-shell degeneracy. Once the training set includes an additional weak-coupling point, the parameter-dependent wavefunction map learns the correct local structure of the solution manifold and extrapolates to the unseen $N=7$ system with the correct $L=\pm 3$ character.

The same conclusion is supported by the angular momentum sector projections shown in Fig.~\ref{fig:N7_U1_density}. These projections provide a direct wavefunction diagnostic beyond visual inspection of the density. For the main text density plots, the inferred wavefunctions place dominant weight in the expected angular momentum sectors. The $N=7$ states are dominated by the paired $L=+3$ and $L=-3$ sectors, the $N=9$ states by the paired $L=+1$ and $L=-1$ sectors, and the closed-shell $N=10$ states by the $L=0$ sector.

\begin{figure}
    \centering
    \includegraphics[width=0.9\linewidth]{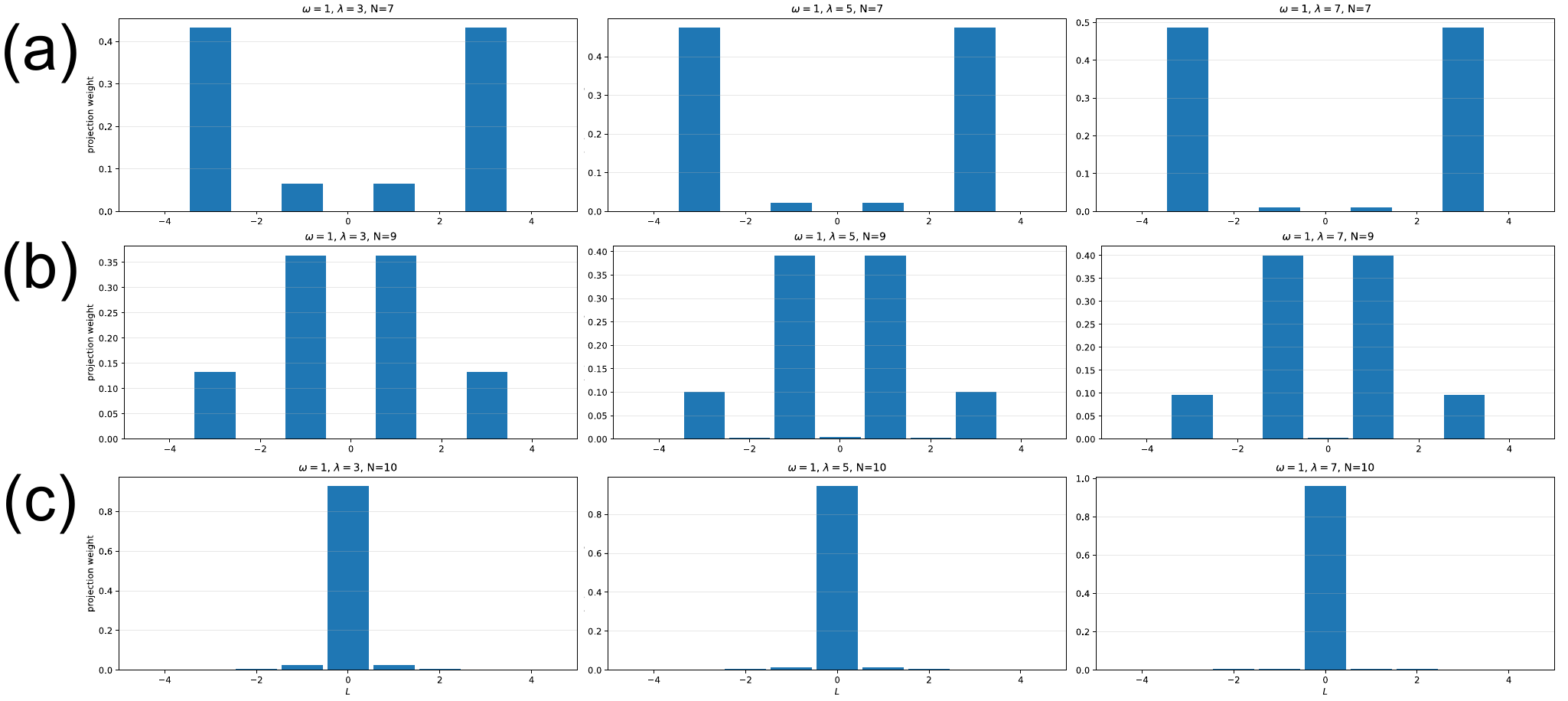}
    \caption{\textbf{Wavefunction angular momentum sector weights for all wavefunctions in Fig.~\ref{fig:densities} in the main text (panels (a)--(c) correspond to $N=7,9,10$, respectively, and the left, middle, and right columns correspond to $\lambda=3.0,5.0,7.0$, respectively).} We project the inferred wavefunction into angular-momentum subspaces. The projection weight is dominated by the $L=3$ sector for $N=7$ (a), $L=1$ sector for $N=9$ (b), and $L=0$ sector for $N=10$ (c), as expected from the modified Hund rules. This provides a direct wavefunction-level diagnostic of the inferred state, beyond visual inspection of the density.}
    \label{fig:N7_U1_density}
\end{figure}

More broadly, this section illustrates how the foundation model should be used. The method learns a continuous map from Hamiltonian parameters and particle number to a variational many-electron wavefunction, but its accuracy depends on whether the training distribution resolves the physical regimes one wants to predict. Open-shell weak-coupling states are demanding because small energetic splittings determine the correct angular momentum sector. The local energy variance provides a quantitative uncertainty measure for such cases, since an exact eigenstate has zero variance,
\begin{equation}
\sigma_E^2
=
\left\langle
\left(E_{\mathrm{loc}}(\mathbf{R})-\langle E_{\mathrm{loc}}\rangle\right)^2
\right\rangle
=0.
\end{equation}
Large or anomalous variance can therefore identify regions where additional variational data are needed. In this example, adding a single weak-coupling training point corrects the angular momentum sector at the unseen $N=7$, $\lambda=1.0$ system. This shows that the foundation model can predict new many-electron solutions accurately when the training set is chosen to resolve the relevant physical structure.

Finally, the angular momentum analysis highlights the flexibility of the approach. Because the output is an explicit many-electron wavefunction, the same variational state can be used to compute any equal-time observable, not only energies and densities. The sector weights are obtained from
\begin{equation}
p_L
=
\frac{
\langle \Psi_\theta | \hat{P}_L | \Psi_\theta\rangle
}{
\langle \Psi_\theta|\Psi_\theta\rangle
},
\qquad
\hat{P}_L
=
\sum_{\alpha\in \mathcal{H}_L}
|\Phi_{L,\alpha}\rangle\langle \Phi_{L,\alpha}|,
\end{equation}
where $\mathcal{H}_L$ is the subspace with total angular momentum $L$. The same procedure can be applied to pair correlations, structure factors, conditional densities, and other equal-time probes of correlated electronic structure.

\section{Optimization stability across system size}
\label{sec:supp_optimization}

Figure~\ref{fig:supp_enfig} compares optimization behavior for a co-trained small-$N$ model (18 parameter groups spanning $N\in \{6,8,10\}$ and $\lambda\in \{0.0,1.0,2.0,5.0,8.0,10.0\}$) and a large-$N$ model ($N=50$, $\lambda\in\{0.0,1.0,2.0,3.0,5.0,6.0,8.0,10.0\}$). The plotted loss is the mean variational energy, averaged over all parameter groups (explicitly shown in Eq.~\ref{eq:multitask_vmc}), and the insets show the corresponding energy variance. Despite training on more systems, the small-$N$ run is markedly stable: the loss drops rapidly and plateaus early, and the variance collapses quickly. The large-$N$ run is noisier and converges more slowly, with occasional spikes in both the energy and variance. This behavior is expected in the large-electron regime: model memory and compute scale roughly as $N^2$ through pairwise features and determinant construction, and with a single 96\,GB GPU, we must constrain batch size and model width. Even with these constraints, the optimization remains well-behaved at long times and the final energies achieve high accuracy, with typical standard errors below $10^{-3}$ on average.

\begin{figure*}
    \centering
    \includegraphics[width=0.9\linewidth]{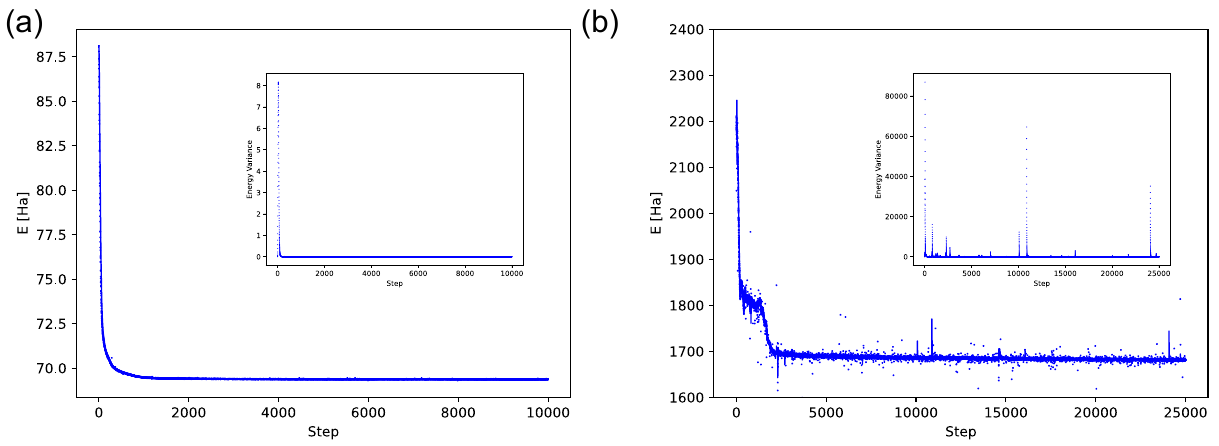}
    \caption{\textbf{Total loss and variance for the small-particle-number run in the main text for (a) $N=6,8,10, \lambda=0.0,1.0,2.0,5.0,8.0,10.0$ and (b) $N=50, \lambda=0.0,1.0,2.0,3.0,5.0,6.0,8.0,10.0$.}}
    \label{fig:supp_enfig}
\end{figure*}

\begin{figure*}
    \centering
    \includegraphics[width=0.9\linewidth]{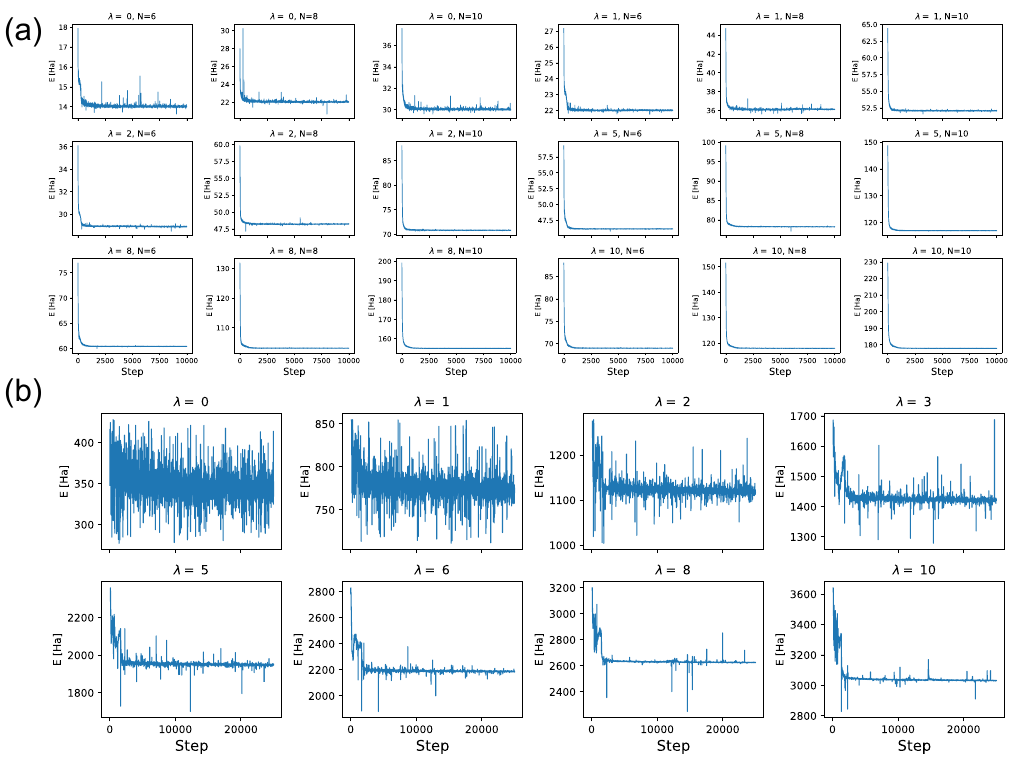}
    \caption{\textbf{Group loss for each parameter group in the main-text runs, corresponding to the runs in Fig.~\ref{fig:supp_enfig}.}}
    \label{fig:supp_enfig_allparams}
\end{figure*}

Figure~\ref{fig:supp_enfig_allparams} resolves the same training into per-parameter curves. For the small-$N$ co-training (top panel), each $(N,\lambda)$ group converges smoothly with only mild transient noise. For the large-$N$ co-training (bottom panel), the relative ordering of noise across $\lambda$ differs from single-task training: lower $\lambda$ runs exhibit visibly larger fluctuations than higher $\lambda$ runs. This does not reflect increased physical difficulty at weak coupling; it is an optimization effect induced by simultaneous training across tasks with widely separated energy scales. When the objective is averaged across parameter groups,
\begin{equation}
\mathcal{L}(\theta)=\frac{1}{G}\sum_{g=1}^{G} E_{\theta}(\mathbf{p}_g),
\qquad
E_{\theta}(\mathbf{p}_g)=
\mathbb{E}_{(\mathbf{R},\mathbf{s})\sim|\Psi_{\theta}(\cdot;\mathbf{p}_g)|^2}\!\left[E_L(\mathbf{R},\mathbf{s};\mathbf{p}_g)\right],
\end{equation}
groups with larger absolute energies (typically larger $\lambda$ and/or larger $N$) contribute gradients with larger magnitude, setting an effective step scale for the shared parameters. For smaller-$\lambda$ groups, this can act like an overly aggressive step size, producing higher apparent noise even when those tasks are individually easy.

This effect can be reduced by balancing the multi-task optimization. One option is to reweight the loss by a $\lambda$-dependent factor or by a running estimate of task scale, e.g., normalizing each group by an energy or variance scale so that gradients are comparable across $\mathbf{p}_g$. Another option is to add a weak regularizer that suppresses excess variance in low-$\lambda$ groups during co-training, or to increase model capacity when resources allow, which reduces competition between tasks and lowers gradient interference. In practice, the observed noise does not prevent accurate convergence: the large-$N$ model still reaches stable final energies and produces highly faithful predicted wavefunctions across the full coupling range, demonstrating that parameter-conditioned co-training remains effective even in the most memory-constrained regime.

\section{Extended data}

Fig.~S6 provides a direct visualization of the model's full many-body wavefunction, $\Psi_\theta(\mathbf{r}_1,\ldots,\mathbf{r}_N)$, which we include to emphasize that the model learns the full many-body wavefunction rather than a reduced derivative quantity. For each system we construct a \emph{conditional wavefunction slice} by fixing $N\!-\!1$ electron coordinates at a representative high-probability configuration (white markers) and scanning the remaining coordinate $\mathbf{r}$ across the 2D plane. The left column shows the conditional probability density $|\psi_{\mathrm{cond}}(\mathbf{r})|^2 \propto |\Psi_\theta(\mathbf{r},\mathbf{r}_2^\star,\ldots,\mathbf{r}_N^\star)|^2$, while the middle column shows the corresponding phase $\phi(\mathbf{r})=\arg \psi_{\mathrm{cond}}(\mathbf{r})$. The sharp boundaries between phase domains coincide with suppressed amplitude and identify nodal contours of the many-body state, information that is inaccessible from one-body densities alone and is not fixed by matching energies. In particular, the presence and topology of these nodal domains demonstrates that the model captures the sign/phase structure (and thus fermionic antisymmetry) of the ground state.

The right column reports the pair correlation function $g(\mathbf{r})$ (conditional density relative to a reference electron, placed at the center of the plot), connecting the wavefunction-level visualization to a standard reduced observable. For the full-shell case $N=10$ at $\lambda=8$ (top row), $g(\mathbf{r})$ exhibits a pronounced exchange-correlation hole at short distance and an approximately isotropic correlation ring, consistent with the rotationally symmetric density profile expected for a filled shell. For the open-shell case $N=7$ at $\lambda=8$ (bottom row), $g(\mathbf{r})$ shows the same correlation hole but with stronger angular modulation and enhanced localization, reflecting the increased role of correlations in an incompletely filled shell. Importantly, these anisotropic features appear in conditional objects ($\psi_{\mathrm{cond}}$ and $g$) while remaining consistent with the rotationally symmetric one-body densities shown in the main text: averaging over particle labels and configurations restores rotational symmetry even when individual conditional slices reveal structured nodal and correlation patterns.

\begin{figure*}
    \centering
    \includegraphics[width=0.9\linewidth]{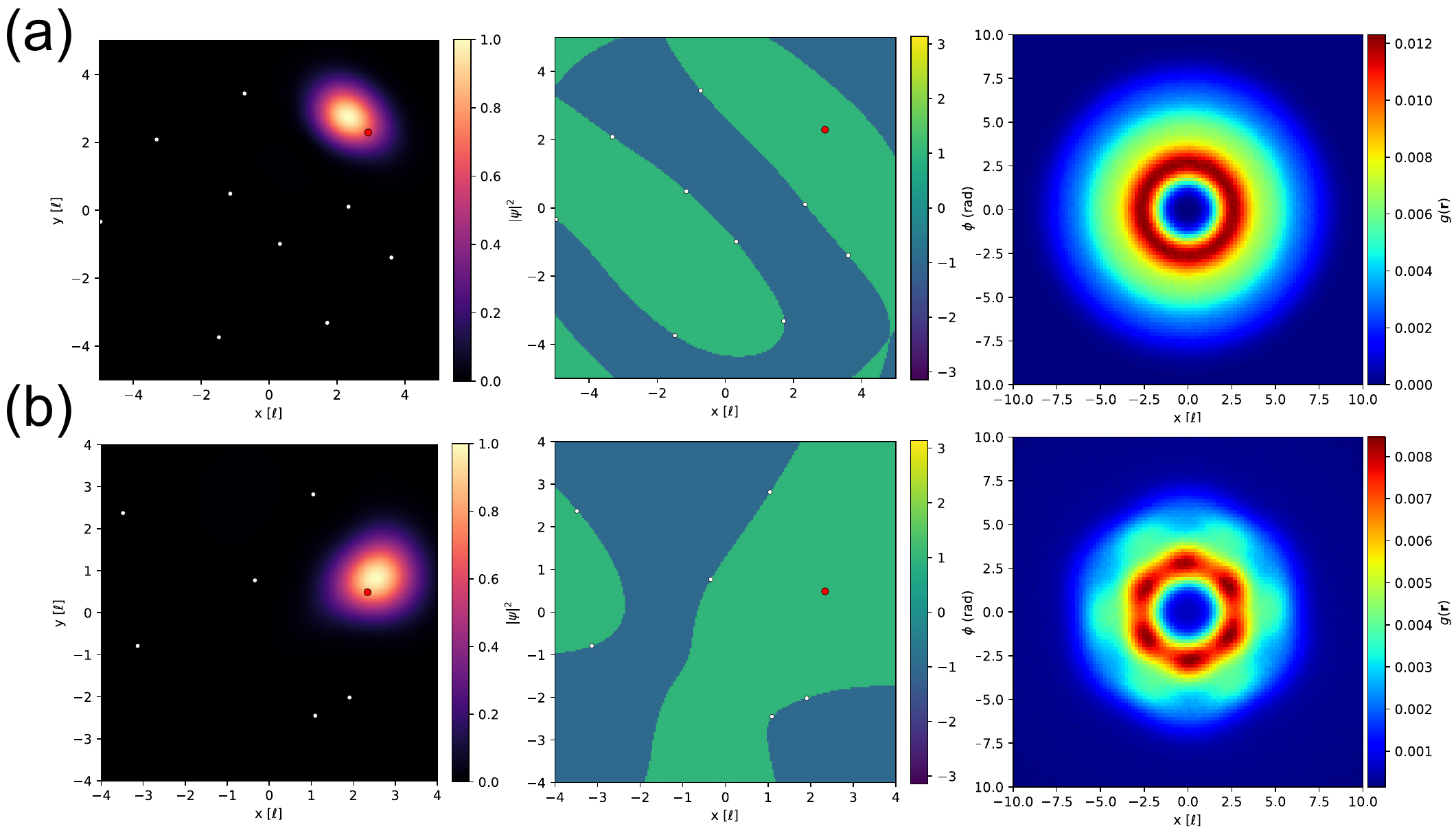}
    \caption{\textbf{Conditional probability (left), phase (middle), and pair correlation plot (right) for (a) $N=10$ and (b) $N=7$ for $\lambda=8.0$.}}
    \label{fig:supp_wavePlots}
\end{figure*}


\end{widetext}